\documentclass[10pt]{article}
\usepackage{amsmath}
\usepackage{times}
\usepackage{graphicx}
\usepackage{color}
\usepackage{multirow}
\usepackage[authoryear]{natbib}
\usepackage{rotating}
\usepackage{latexsym}
\usepackage[nolist,nohyperlinks]{acronym} % Acronyms
\newcommand{\captionfonts}{\normalsize}

\makeatletter  
\long\def\@makecaption#1#2{%
  \vskip\abovecaptionskip
  \sbox\@tempboxa{{\captionfonts #1: #2}}%
  \ifdim \wd\@tempboxa >\hsize
    {\captionfonts #1: #2\par}
  \else
    \hbox to\hsize{\hfil\box\@tempboxa\hfil}%
  \fi
  \vskip\belowcaptionskip}
\makeatother   
\begin{document}

\begin{acronym}[ticktocktick]
	\acro{MEA}{microelectrode array}	
	\acro{div}{days \textit{in vitro}}
	\acro{SC}{Spike-contrast}
	\acro{BIC}{bicuculline}
	\acro{STTC}{spike time tiling coefficient}
	\acro{CC}{cross-correlation}
	\acro{PS}{phase synchronization}
	\acro{MI}{mutual information}
	\acro{ISI}{interspike interval}
	\acro{NBQX}{2,3-dioxo-6-nitro-l,2,3,4-tetrahydrobenzoquinoxaline-7-sulphonamide}
	\acro{GABA$_A$}{$\gamma$-aminobutyric acid}
	\acro{TDNS}{total deviation of the normalized synchrony}
\end{acronym}

\ \vspace{20mm}\\

This is the author’s final version. The article has been accepted for publication in Neural Computation (Volume 32, Issue 5).

\hspace{13.9cm}

{\LARGE{Comparison of different spike train synchrony measures regarding their robustness to erroneous data from bicuculline induced epileptiform activity}}

\ \\
{\bf \large Manuel Ciba$^{\displaystyle 1, \displaystyle *}$, 
Robert Bestel$^{\displaystyle 1}$,
Christoph Nick$^{\displaystyle 1}$,
Guilherme Ferraz de Arruda$^{\displaystyle 2}$,
Thomas Peron$^{\displaystyle 3}$,
Comin C\'{e}sar Henrique$^{\displaystyle 4}$,
Luciano da Fontoura Costa$^{\displaystyle 5}$,
Francisco Aparecido Rodrigues$^{\displaystyle 3}$,
Christiane Thielemann$^{\displaystyle 1}$
}\\

{$^{\displaystyle 1}$biomems lab, University of Applied Sciences Aschaffenburg, 63743 Aschaffenburg, Germany.}\\
{$^{\displaystyle 2}$ISI Foundation, Via Chisola 5, 10126 Torino, Italy.}\\
{$^{\displaystyle 3}$Institute of Mathematics and Computer Science, University of S\~{a}o Paulo - S\~{a}o Carlos, SP 13566-590, Brazil.}\\
{$^{\displaystyle 4}$Department of Computer Science, Federal University of S\~ao Carlos - S\~ao Carlos, SP, Brazil.}\\
{$^{\displaystyle 5}$Instituto de F\'{\i}sica de S\~{a}o Carlos, Universidade de S\~{a}o Paulo, S\~{a}o Carlos, S\~ao Paulo, Brazil.}\\
{$^{\displaystyle *}$Corresponding Author, manuel.ciba@th-ab.de} \\

%\ \\[-2mm]

\thispagestyle{empty}
\markboth{}{NC instructions}
\ \vspace{-0mm}\\

%\newpage

%
%Abstract
\begin{center} {\bf Abstract} \end{center}

As synchronized activity is associated with basic brain functions and pathological states, spike train synchrony has become an important measure to analyze experimental neuronal data. 
Many different measures of spike train synchrony have been proposed, but there is no gold standard allowing for comparison of results between different experiments.
This work aims to provide guidance on which synchrony measure is best suitable to quantify the effect of epileptiform inducing substances (e.g. \ac{BIC}) in \textit{in vitro} neuronal spike train data.

Spike train data from recordings are likely to suffer from erroneous spike detection, such as missed spikes (false negative) or noise (false positive). Therefore, different time-scale dependent (\acl{CC}, \acl{MI}, \acl{STTC}) and time-scale independent (Spike-contrast, \acl{PS}, A-SPIKE-synchronization, A-ISI-distance, ARI-SPIKE-distance) synchrony measures were compared in terms of their robustness to erroneous spike trains.

For this purpose, erroneous spike trains were generated by randomly adding (=false positive) or deleting (=false negative) spikes (=\textit{in silico} manipulated data) from experimental data. In addition, experimental data were analyzed using different spike detection threshold factors in order to confirm the robustness of the synchrony measures. All experimental data were recorded from cortical neuronal networks on \ac{MEA} chips, which show epileptiform activity induced by the substance \ac{BIC}. 

As a result of the \textit{in silico} manipulated data, Spike-contrast was the only measure being robust to false negative as well as false positive spikes. Analyzing the experimental data set revealed that all measures were able to capture the effect of \ac{BIC} in a statistically significant way, with Spike-contrast showing the highest statistical significance even at low spike detection thresholds.  
In summary, we suggest the usage of Spike-contrast to complement established synchrony measures, as it is time-scale independent and robust to erroneous spike trains.
\newpage

\section{Introduction}
Synchrony is generally accepted to be an important feature of basic brain functions \citep{engel2001dynamic, Ward2003Synchronous, rosenbaum2014correlated} and pathological states \citep{Pare1990Neuronal,Fisher2005Epileptic,Truccolo2014neuronal,Arnulfo2015Characterization}.
Measuring synchrony between neural spike trains is a common method to analyze experimental data e.g. recordings from \textit{in vitro} neuronal cell cultures with microelectrode arrays (MEA) \citep{Selinger2004Measuring, chiappalone2006dissociated, Chiappalone2007Network, Eisenman2015Quantification, Flachs2016cell} or from \textit{in vivo} experiments \citep{li2011acute}. 
As an example for \textit{in vitro} neuronal cell cultures on \ac{MEA} chips, \citet{Sokal2000} reported that synchrony reliably increased due to the substance \acf{BIC}, while the usual applied quantification method ``spike rate'' increased or decreased.
In order to quantify synchrony, many spike train synchrony measures have been proposed based on different approaches. Some of them belong to the class of time-scale dependent measures. This means that at the beginning of the analysis the user has to select the desired time scale (e.g. bin size) \citep{Selinger2004Measuring, Cutts2014Detecting}. The second class contains time-scale independent measures, which automatically adapt their time-scale parameter according to the data \citep{Satuvuori2017Measures, Ciba2018Spike}. However, there is no gold standard for the evaluation of synchrony in experimental data. This is because there is no common definition of synchrony between spike trains. To be more specific, each synchrony measure can be considered as its own definition of synchrony, extracting different features from the data. 
This situation is unsatisfactory as data interpretations are not comparable. Therefore, a guidance would be desirable on which synchrony measure to use for specific data. 

When it comes to the analysis of experimental spike train data, the data are likely to suffer from erroneous spike detection. For example, spikes are missed as they are buried in noise (false negative) or noise is misinterpreted as spikes (false positive). At low signal-to-noise ratios (SNR), even advanced spike detection methods are affected by missed or misinterpreted spikes \citep{lieb2017stationary}.

Hence, a synchrony measure that operates on spike trains from experimental data should be as robust as possible to such erroneous spike trains. \\ 

In order to approach a guidance to analyze epileptiform spike trains from \textit{in vitro} neuronal networks, the performance of different synchrony measures was compared with the focus on robustness to erroneous spike trains. 
Well known time-scale dependent measures, like \acf{CC}, \acf{MI}, and \acf{STTC} and time-scale independent measures, like Spike-contrast, \acf{PS}, A-SPIKE-synchronization, A-ISI-distance, A-SPIKE-distance, and ARI-SPIKE-distance were applied to two types of data sets: (1) \textit{in silico} manipulated data and (2) experimental data.

(1) The in silico manipulated data are based on the experimental data and were used to simulate erroneous spike train data
by randomly adding spikes (false positive) or deleting spikes (false negative). As a requirement, the synchrony measures should be robust to added and deleted spikes.

(2) The experimental data were recorded from primary cortical networks grown \textit{in vitro} on \ac{MEA} chips. Neuronal networks were exposed to the \ac{GABA$_A$} receptor antagonist \acf{BIC} in order to increase the synchrony level of the network activity. Spike detection threshold factor was varied in order to vary the level of false positive and false negative spikes. The synchrony measures were tested for their ability to find significant synchrony changes induced by \ac{BIC}.

%%%%%%%%%%%%%%%%%%%%%%%%%%%%%%%%%%%%%%%
% METHODS
%%%%%%%%%%%%%%%%%%%%%%%%%%%%%%%%%%%%%%%
\section{Material and methods} \label{Section_MaterialAndMethods}

\subsection{Synchrony measures} \label{Section_SynchronyMeasures}
In this section the synchrony measures used in this study are briefly described. To consider a wide range of synchronization measures, a representative group of linear and nonlinear methods as well as time-scale dependent and independent methods were chosen.
For a detailed definition see the respective original publication. Since there is no specific publication on how to apply \ac{MI} and \ac{PS} to spike train data, their definitions are provided in appendix~\ref{Appendix1}.  

\paragraph{Time-scale dependent:}

\begin{itemize}

\item \textbf{Cross-correlation (CC)} based methods are probably most popular to measure synchrony \citep{Cutts2014Detecting}. Here we use a definition by \citet{Selinger2004Measuring} that was specially proposed for \textit{in vitro} experiments and had also been used by \citet{chiappalone2006dissociated}. According to the definition, synchrony between two spike trains is measured by binning the spike trains into a binary signal and then calculating the cross-correlation without shifting the signals. \citet{Selinger2004Measuring} proposed a bin size of 500~ms and was able to detect synchrony changes in spinal-cord cultures mediated by the chemicals \ac{BIC}, strychnin, and \ac{NBQX}. Due to the bin size parameter, \ac{CC} is time-scale dependent. A bin size of 500~ms is also used in this study (see Section~\ref{sec:ParameterChoice}).
   
\item \textbf{Mutual information (MI)} is a measure from the field of information theory and is -- in contrast to \ac{CC} -- able to capture non-linear dependencies. In this work, \ac{MI} measures the synchrony between two spike trains by binning the spike trains into binary signals and quantifying the redundant information \citep{Cover012}. Therefore, this version of \ac{MI} is time-scale dependent using a bin size of 500~ms (see Section~\ref{sec:ParameterChoice}).

\item \textbf{Spike time tiling coefficient (STTC)} measures the synchrony between two spike trains  and has been proposed by \citet{Cutts2014Detecting} as a spike rate independent replacement of a synchrony measure called "correlation index" by \citet{wong1993transient}. Reanalysis of a study of retinal waves using \ac{STTC} instead of the "correlation index" significantly changed the result and conclusion \citep{Cutts2014Detecting}. \ac{STTC} is a time-scale dependent measures as it needs a predefined time window $\Delta t$ in which spikes are considered synchronous. Referring to the work of \citet{Cutts2014Detecting}, a time window of 100~ms was used in this work (see Section~\ref{sec:ParameterChoice}).

\end{itemize}

\paragraph{Time-scale independent:}

\begin{itemize}

\item \textbf{Phase synchronization (PS)} measures the synchrony between spike trains in two steps. First step is to assign a linear phase procession from $0$ and $2\pi$ to every \ac{ISI}. Second step is quantifying the common phase evolution of all spike trains via an order parameter defined by \citet{Pikovsky03}. \ac{PS} is time-scale independent and -- to the best of our knowledge -- has not been systematically compared with other measurements in studies of spike train synchrony yet or has even never been used to measure synchrony of neural spike trains.

\item \textbf{Spike-contrast} is a time-scale independent synchrony measure based on the temporal ``contrast'' of the spike raster plot (activity vs. non-activity in certain temporal bins) and not only provides a single synchrony value, but also a synchrony curve as a function of the bin size, or in other words, as a function of the time-scale \citep{Ciba2018Spike}. Here, instead of the synchrony curve only the single synchrony value was used.

\item \textbf{A-SPIKE-synchronization} is a time-scale independent and parameter free coincidence detector \citep{Satuvuori2017Measures}. It measures the similarity between spike trains and is the adaptive generalization of SPIKE-synchronization \citep{kreuz2015spiky}. In the adaptive versions a decision is made if the spike trains are compared considering their local or global time-scale, which is advantageous for data containing different time-scales like regular spiking and bursts. 

\item \textbf{A-ISI-distance} is a time-scale independent and parameter free distance measure \citep{Satuvuori2017Measures}. It measures the instantaneous rate difference between spike trains and is the adaptive generalization of ISI-distance \citep{Kreuz2007Measuringa}.

\item \textbf{A-SPIKE-distance} is a time-scale independent and parameter free distance measure \citep{Satuvuori2017Measures}. It measures the accuracy of spike times between spike trains relative to local firing rates and is the adaptive generalization of SPIKE-distance \citep{kreuz2011time, Kreuz2013Monitoring}. 

\item \textbf{ARI-SPIKE-distance} is the rate independent version of A-SPIKE-distance \citep{Satuvuori2017Measures}. It measures the accuracy of spike times between spike trains without using the relative local firing rate. Some of the original version have already been applied to experimental neuronal data. For example \citet{andrzejak2014detecting} used ISI-distance and SPIKE-distance and \citet{dura2016restoring} used SPIKE-distance and SPIKE-synchronization. \citet{espinal2016quadrupedal} applied SPIKE-distance to simulated data.   
 
\end{itemize}

In order to get a final synchrony value over all recorded spike trains, synchrony between all spike train pairs was calculated and averaged. With the exception of Spike-contrast, which already yields a single synchrony value between all spike trains due to its multivariate nature.  \newline

Note that all synchrony measures are designed to provide a value between 0 (miminum synchrony) and 1 (maximum synchrony). Only \ac{CC} and \ac{STTC} are able to yield negative values in case of anticorrelation. The distance measures A-ISI-distance, A-SPIKE-distance, and ARI-SPIKE-distance naturally provide values between 0 (minimal distance or maximum synchrony) and 1 (maximum distance or minimum synchrony). Therefore, their values were substracted from 1 to make the distance measures comparable to the synchrony measures. The MATLAB\textsuperscript{\textregistered} (MATLAB 2016a, MathWorks, Inc., Natick, Massachusetts, USA) source code of A-SPIKE-synchronization, A-ISI-distance, A-SPIKE-distance, and ARI-SPIKE-distance was downloaded along with the tool called \textit{cSPIKE}\footnote{\normalsize http://wwwold.fi.isc.cnr.it/users/thomas.kreuz/Source-Code/cSPIKE.html}. The Spike-contrast\footnote{\normalsize https://github.com/biomemsLAB/Spike-Contrast} and \ac{MI}\footnote{\normalsize https://de.mathworks.com/matlabcentral/fileexchange/28694-mutual-information} MATLAB source code also was taken from online sources. \ac{STTC} python code\footnote{\normalsize http://neuralensemble.org/elephant/} was translated into MATLAB code. MATLAB code for \ac{PS} was specifically programmed for this work. All MATLAB functions and scripts used for this work are provided online\footnote{\normalsize https://github.com/biomemsLAB/SynchronyMeasures-Robustness}.

\subsection{\textit{In silico} manipulated data} \label{Section_HybridTestData}

\begin{figure}[h]
%\nopagenumber
%\renewcommand{\baselinestretch}{1.0}
\begin{center}
\includegraphics[width=1.0\textwidth]{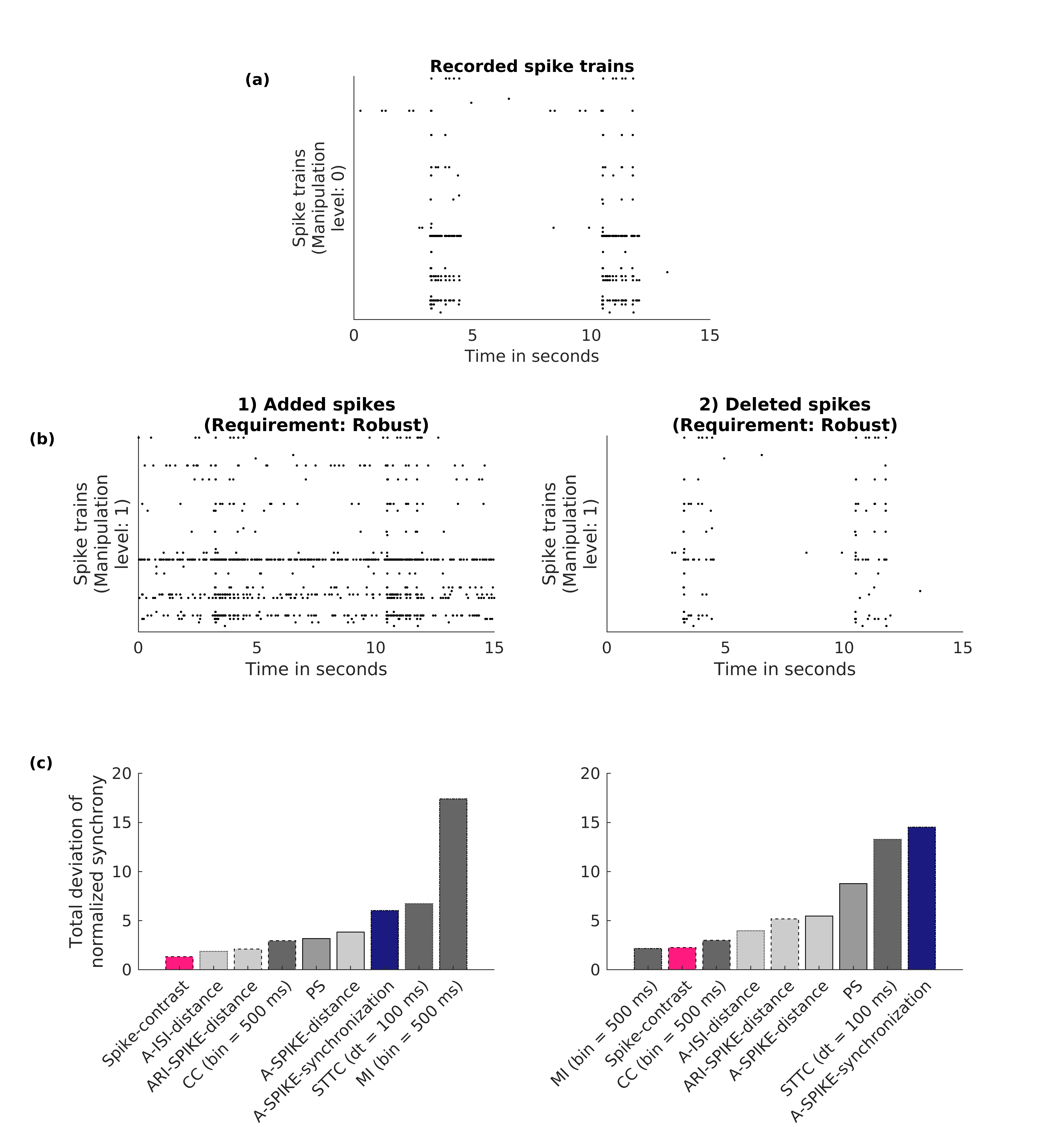}
\end{center}
\caption[]{\textbf{Comparison of different synchrony measures regarding their robustness to erroneous spike trains.} \textbf{(a)} Spike trains recorded from 60 electrodes with 10~$\mu$M \ac{BIC}. Each line represents the spike train of one electrode (only first 15~s of 300~s are displayed). \textbf{(b)} Same as (a) but with maximum manipulation level applied (left: Simulation of false positive spikes, right: Simulation of false negative spikes). In total 10 different recordings were used to generate the \textit{in silico} manipulated data (for details see Sec.~\ref{Section_HybridTestData}). \textbf{(c)} Sum of synchrony deviation over all manipulation levels (denoted as \acf{TDNS}, see Eq.~\ref{eq:TDNS}). The lower the \ac{TDNS} value of a synchrony measure, the more robust it is to spike train manipulation.}    
\label{Fig:Result_Hybrid}
\end{figure}

Two sets of \textit{in silico} manipulated data were generated featuring added spikes (=false positive spikes) and deleted spikes (=false negative spikes). As the measures \ac{CC}, \ac{MI}, and \ac{STTC} are time-scale dependent, the \textit{in silico} manipulated data are based on the experimental data (see Section~\ref{Section_ExpTestData}) in order to obtain realistic time-scales. In total, 10 recordings from 5 independent networks (N = 5) were used (5 without and 5 with 10~$\mu$M \ac{BIC}). Each recording had a length of 300~s and up to 60 active electrodes. The following procedures were applied for every active electrode (active if at least 6 spikes per minute, see Section \ref{Section_Spikedetection}) with $X$ being the spike train of the original electrode and $Y$ being the manipulated spike train:

\begin{itemize}

\item[1)] \textbf{Added spikes:} 
Spike train $Y$ was generated by copying spike train $X$ and adding $N_{add}$ spikes to $Y$ with temporal positions randomly assigned in the range of $(0, 300]$ seconds. In case of identical spike times, new random spike times were generated until all spike times were unique. Depending on the manipulation level the number of added spikes was
\begin{equation}
N_{add} = L \cdot 0.1 \cdot N_X, \label{eq:AddedSpikes}
\end{equation}
with $N_X$ being the number of spikes in spike train $X$ and $L$ being the manipulation level in the range of $L = [0, 0.1, 0.2 ...  1]$ ($L = 0$: No manipulation, $L = 1$: 10\% random spikes were added).
For each $L$, 40 independent random manipulations ($n_{manipulated}=40$) were performed (see Fig.~\ref{Fig:Result_Hybrid}~(b~2) for example spike trains).

\item[2)] \textbf{Deleted spikes:}
Spike train $Y$ was generated by copying spike train $X$ and deleting $N_{delete}$ randomly selected spikes from $Y$. Depending on the manipulation level the number of deleted spikes was
\begin{equation}
N_{delete} = L \cdot 0.9 \cdot N_X, \label{eq:DeletedSpikes}
\end{equation}
where $N_X$ is the number of spikes in spike train $X$ and $L$ is the manipulation level in the range of $L = [0, 0.1, 0.2, ...  1]$ ($L = 0$: No manipulation, $L = 1$: 90\% of all spikes were deleted). Note that the maximum range was restricted to 90\% of $N_X$ as some of the tested synchrony measures were not defined for empty spike trains. For each $L$, 40 independent random manipulations ($n_{manipulated}=40$) were performed (see Fig.~\ref{Fig:Result_Hybrid}~(b~3) for example spike trains).

\end{itemize}

A single synchrony value was then calculated for every recording ($N=10$) and every random manipulation ($n_{manipulated}=40$). Overall, for every manipulation level $400$ synchrony values were calculated per synchrony measure.   

As some of the synchrony measures differ in terms of their minimum value for Poisson spike trains with equal rate (e.g. $0.295$ for SPIKE-distance and $0.5$ for ISI-distance \citep{Kreuz2013Monitoring}), they could not be compared directly. 
Therefore, synchrony values of each measure were rescaled to their lowest possible synchrony value defined as synchrony of a data set made of Poisson spike trains $Y_{L,random}$ at the manipulation level $L$. The number of spikes in $Y_{L,random}$ was equal to the number of spikes in $Y_L$ (being the manipulated spike train $Y$ at the manipulation level $L$), to account for the spike rate dependence of some synchrony measures, as reported in \citet{Cutts2014Detecting}. The scaling was done with

\begin{equation}
s'_L = \frac{s_L - \bar{s}_{L,random}}{1 - \bar{s}_{L,random}},
\end{equation}

where $s_L$ is the synchrony value at a manipulation level $L$, $\bar{s}_{L,random}$ is the mean synchrony value of the data set consisting of Poisson spike trains $Y_{L,random}$, and value ``1 '' representing the largest possible synchrony value (all synchrony measures were able to yield ``1 '' for identical spike trains). As ten different recordings with different synchrony level were used, all synchrony values were normalized, allowing to calculate the mean over all recordings and all random realizations. The normalization was done with

\begin{equation}
s''_{L} = \frac{s'_L}{s'_{L=0}},
\end{equation}

where $s'_{L=0}$ is the synchrony of the orginal spike train (manipulation level $L = 0$), and $s''_{L}$ the normalized synchrony value. All normalized synchrony values of all recordings and all random realizations taken together are denoted as $s''_{L, all}$.
In order to ease the comparison between the different synchrony measures, the \ac{TDNS} over the manipulation level was calculated as

\begin{equation} 
TDNS = \sum_{L=0}^{1} std(s''_{L, all}), \label{eq:TDNS}
\end{equation}

where $std()$ is the standard deviation. The lower the \ac{TDNS}, the more robust the synchrony measure against the spike train manipulation procedure. 
 
\subsection{Experimental data} \label{Section_ExpTestData}

\subsubsection{Cell culture and electrophysiological recordings}
Experimental data used for this study were recorded from primary cortical neurons (Lonza Ltd, Basel, Switzerland) harvested from embryonic rats (E18 and E19). Cell cultivation followed a modified protocol based on \citet{Otto03}. Briefly, vials containing 4$\times$10$^6$ cells were stored in liquid nitrogen at -196$^0$C. After thawing, cells were diluted drop-wise with pre-warmed cell culture medium and seeded at a density of 5000\,cells/mm$^2$ onto Poly-D-Lysine and Laminin coated (Sigma Aldrich Co. LLC, St. Louis, USA) microelectrode arrays (60MEA200/30iR-Ti, Multichannel Systems MCS GmbH, Reutlingen, Germany). Twice a week, half of the medium was replaced with fresh, pre-warmed medium. \newline
Neuronal signals were recorded extracellularly at a sampling rate of 10\,kHz outside the incubator at 37$^0$C employing a temperature-controller (Multichannel Systems MCS GmbH).
For drug-induced increase of network synchronization, \acf{BIC} (Sigma Aldrich Co. LLC) was applied to the neuronal cell culture after 21~\ac{div} at a concentration of 10\,$\mu$M. 
\ac{BIC} is a competitive antagonist of the GABA${_A}$ receptor. Since it blocks the inhibitory function of GABA${_A}$ receptors, its application yields an increased incidence of synchronized burst events \citep{Jungblut09}, see Fig.~\ref{Fig:spikegram} (b) and (c). Neuronal activity was recorded for 5 minutes before and 5 minutes during the application of \ac{BIC}.

\subsubsection{Spike detection} \label{Section_Spikedetection}

\begin{figure}[h]
	\begin{center}
	\includegraphics[width=.8\textwidth]{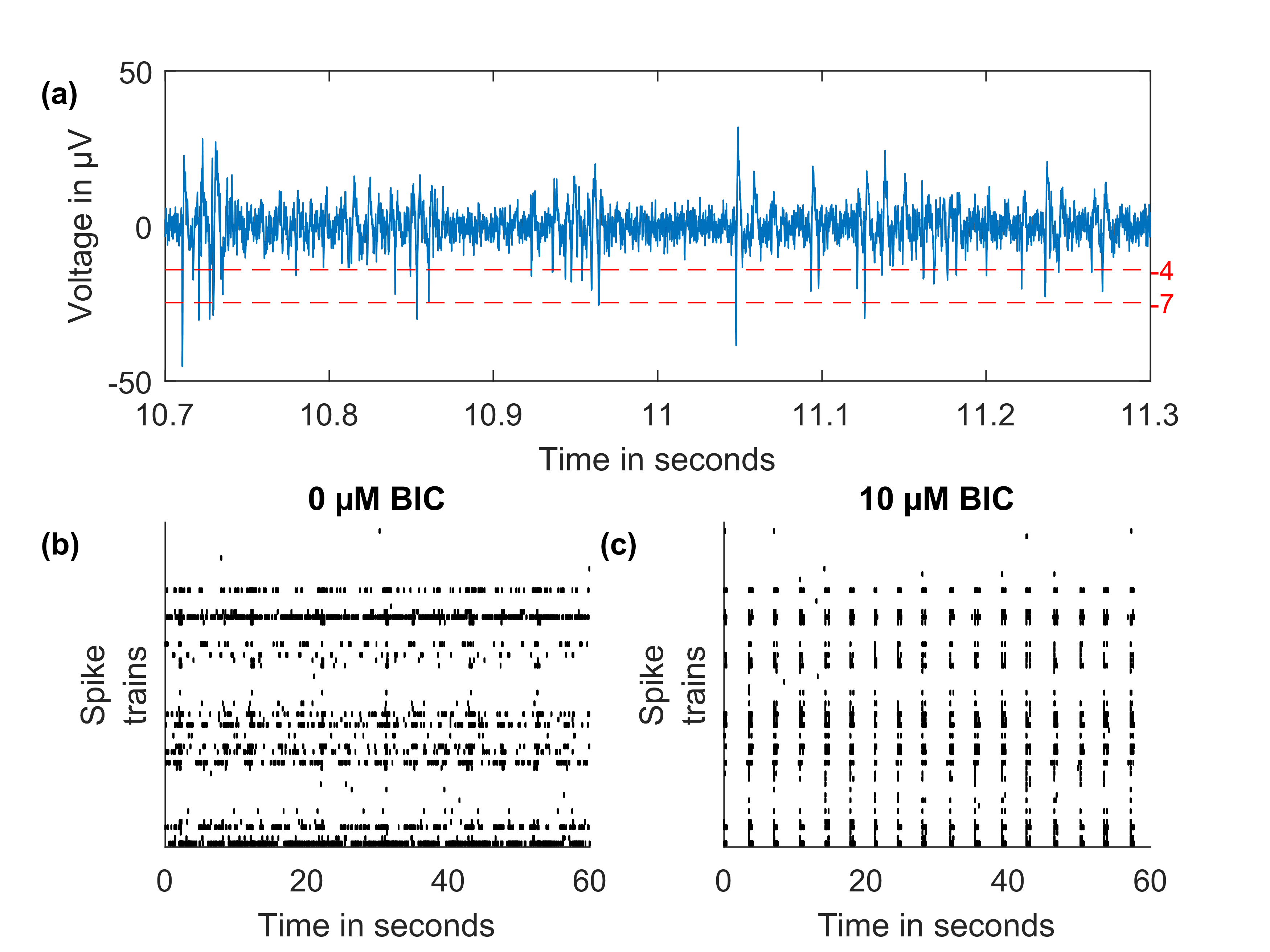}
	\end{center}
	\caption{{\bf Exemplary illustration of the experimental data.} \textbf{(a)} Signal recorded from one \ac{MEA} electrode (50~Hz high-pass filtered). Thresholds are displayed for the lowest (4) and highest (7) threshold factors used for spike detection. The higher the threshold factor, the less spikes are detected. \textbf{(b)} Rasterplot of the spontaneous activity of one \ac{MEA} chip without \ac{BIC} (only first 60~s of 300~s are shown). \textbf{(c)} Same as (b) but with 10~$\mu$M \ac{BIC}. Spontaneous activity was recorded with a 60 electrode \ac{MEA} chip at 23~\ac{div}. Compared to the native state, the administration of 10~$\mu$M \ac{BIC} increased the level of synchrony.}
	\label{Fig:spikegram}
\end{figure}

Raw data were stored for offline spike detection with \textit{DrCell}, a custom-made Matlab (MathWorks Inc., Natick, Massachusetts, U.S.A.) software tool, developed by \citet{Nick013}. After applying a high-pass filter with a cutoff frequency of 50\,Hz, spikes were separated from noise using a threshold-based algorithm, with a threshold calculated by multiplying the standard deviation of the noise by a factor of 5 (or more precisely -5 as only negative spikes were used for analysis). As soon as the raw signal underran the threshold, the time of the minimum voltage value was defined as a spike time stamp (Fig.~\ref{Fig:spikegram}~(a)). Additionally, in order to alter the level of false positive and false negative spikes, spike detection was also conducted using threshold factors of 4, 6, and 7. 
For subsequent spike train analysis, only electrodes with more than 5 spikes per minute were considered active and used for analysis \citep{Novellino2011Development}.

\subsubsection{Statistical analysis} \label{sec:Statistics}
Since it is known that \ac{BIC} causes an increase of network synchrony in cortical neurons \textit{in vitro} \citep{Sokal2000,Chiappalone07,Eisenman2015Quantification}, a one-tailed statistical test was applied. More specifically, a paired t-test was applied under the null hypothesis that \ac{BIC} does not increase synchrony. The lower the p-value, the lower the probability that there is no synchrony change, and hence the better the synchrony measure's sensitivity to \ac{BIC}. 

\subsection{Parameter choice} \label{sec:ParameterChoice}

\begin{figure}[h]
%\nopagenumber
%\renewcommand{\baselinestretch}{1.0}
\hfill
\begin{center}
\includegraphics[width=.5\textwidth]{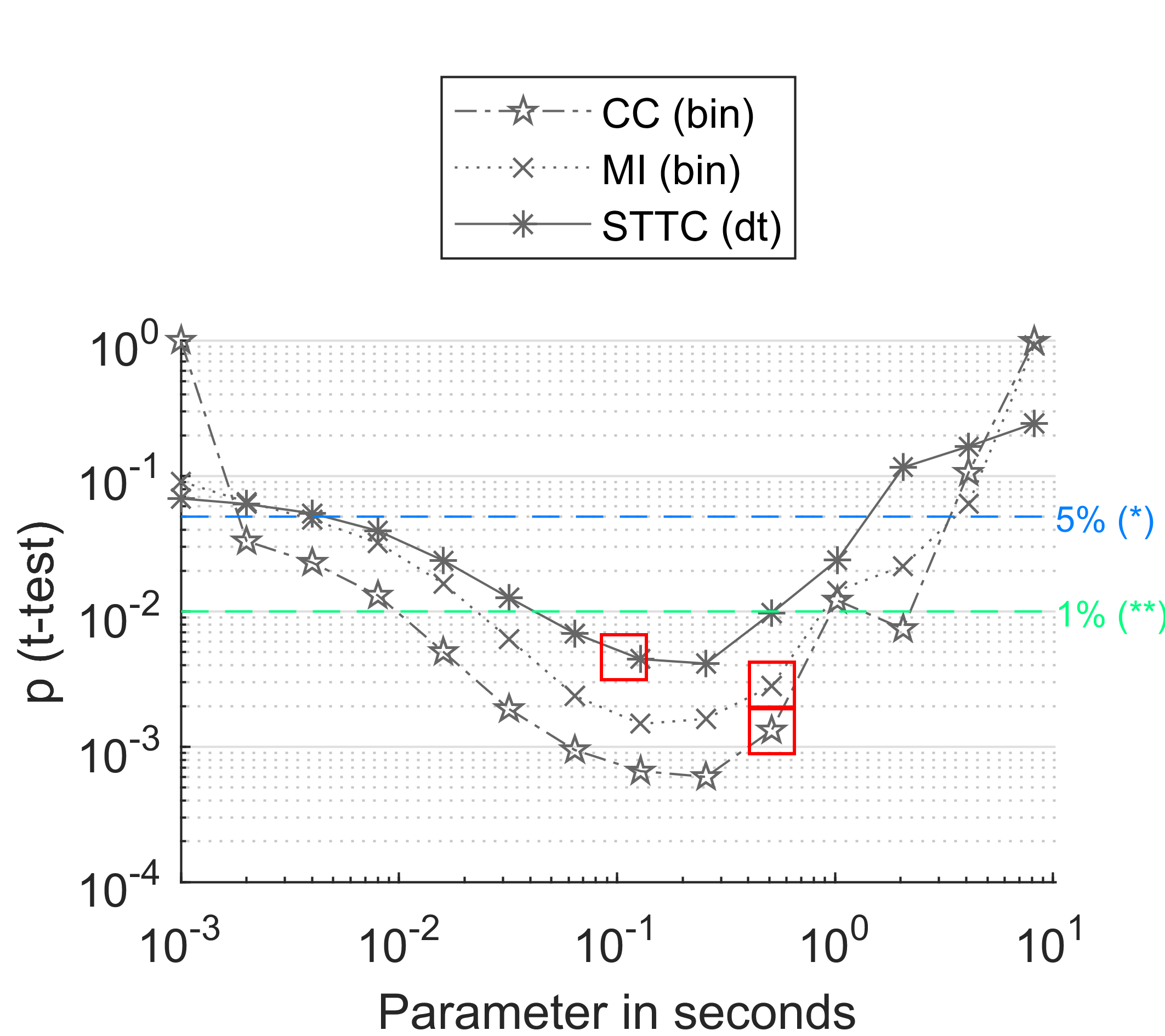}
\end{center}
\caption[]{\textbf{Influence of the parameter choice on the statistical significance of the experimental data.} The synchrony measures \ac{CC}, \ac{MI}, and \ac{STTC} depend on a parameter which has to be chosen by the user. For each synchrony measure, the experimental data described in Section~\ref{Section_ExpTestData} were analyzed using different parameter values. After that, a t-test was applied to the data with the null hypothesis that synchrony values of the data with \ac{BIC} are lower or equal to synchrony values of the data without \ac{BIC}. In this work, a bin size of 500~ms was used for \ac{CC} and a $dt$ of 100~ms for \ac{STTC} (red squares), as these values were suggested in their original publications. For \ac{MI}, the same bin size as for \ac{CC} was used due to the identical binning procedure. 
}
\label{Fig:ParameterChoice}
\end{figure}

As the synchrony measures \ac{CC}, \ac{MI}, and \ac{STTC} depend on a time-scale parameter that directly influences the synchrony definition, an appropriate parameter value had to be chosen. Therefore, we analyzed the experimental \ac{BIC} data using different parameter values and conducted a statistical hypothesis test. The lower the p-value, the higher the probability that an increase in synchrony occurred.
Parameter values that lead to low p-values are therefore desirable (given that \ac{BIC} results in increased level of synchrony). \ac{CC} proposed by \citet{Selinger2004Measuring} chose a bin size of 500~ms, which also gave reasonable low p-values in our analysis (Fig.~\ref{Fig:ParameterChoice}). Even if smaller bin sizes yielded lower p-values in our data, we stuck to the predefined 500~ms. For \ac{MI}, we also chose a bin size of 500~ms as for smaller bin sizes the p-value only improved slightly. The parameter $dt$ of \ac{STTC} was set to 100~ms according to the original publication from \citet{Cutts2014Detecting}. Note that in our cortical data a larger dt value would only slightly improve the p-value.

%%%%%%%%%%%%%%%%%%%%%%%%%%%%%%%%%%%%%%%
% Results and Discussion
%%%%%%%%%%%%%%%%%%%%%%%%%%%%%%%%%%%%%%%
\section{Results}

\subsection{Comparison of synchrony measures for \textit{in silico} manipulated data} \label{sec:simulation}

\begin{itemize}
\item[1)] \textbf{Added spikes:} The robustness of measures was studied by randomly adding false positive spikes to the basic signals. For a robust synchrony measure low sensitivity to false positive spikes is desirable as such noise may falsify the results. 
As displayed in Fig.~\ref{Fig:Result_Hybrid}~(e~1) Spike-contrast was most robust to false positive spikes indicated by the small \ac{TDNS} value of around $1$, closely followed by A-ISI-distance and ARI-SPIKE-distance. \ac{MI} showed by far the largest synchrony deviation indicated by a \ac{TDNS} value of around $17$. \newline

\item[2)] \textbf{Deleted spikes:}  A similar picture occurs after increasingly deleting spikes in order to simulate false negative spikes, as robustness to false negative spikes was required. In Fig.~\ref{Fig:Result_Hybrid}~(e~2) results of synchrony measure dependency to false negative spikes are shown. Here, \ac{MI} and Spike-contrast were most robust to false negative spikes yielding the smallest \ac{TDNS} value of around $2$. \ac{STTC} and A-SPIKE-synchronization showed largest \ac{TDNS} value of around $14$. \newline

\end{itemize}

\subsection{Comparison of synchrony measures for experimental data}

\begin{figure}[h]
%\nopagenumber
%\renewcommand{\baselinestretch}{1.0}
\hfill
\begin{center}
\includegraphics[width=.6\textwidth]{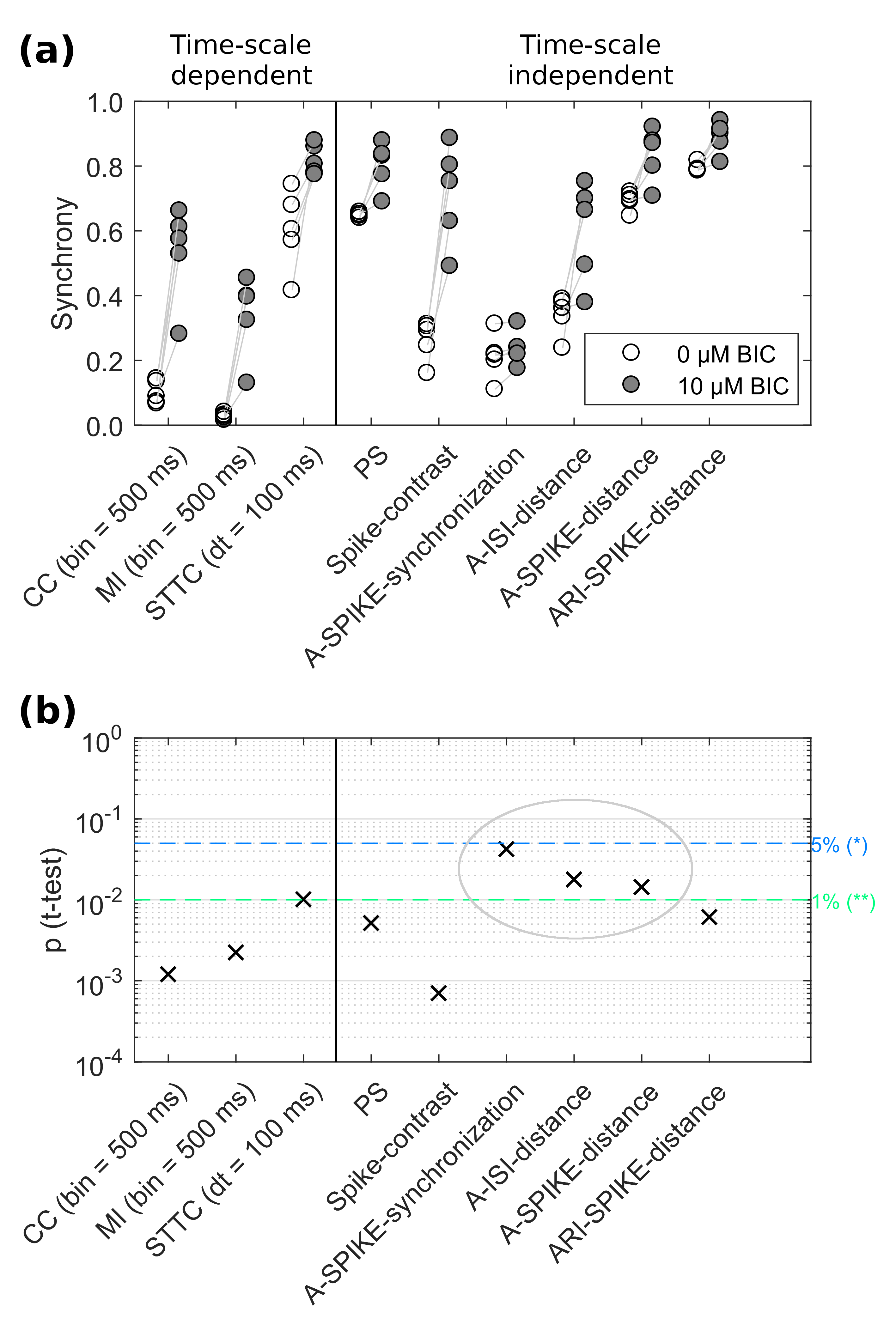}
\end{center}
\caption[]{\textbf{Results of the application of different time-scale dependent and time-scale independent synchrony measures to a set of experimental data.} Experimental data were recorded from cortical cell cultures on \ac{MEA} chips ($N=5$) in the absence and presence of 10\,$\mu$M \ac{BIC}. \textbf{(a)} Absolute synchrony values of each synchrony measure for data without \ac{BIC} (white circles) and with \ac{BIC} (gray circles). \textbf{(b)} P-values for each synchrony measure from a one-tailed paired t-test assuming an increase in synchrony.  The lower the p-value, the better the synchrony measure's ability to capture the effect of \ac{BIC}. Grey ellipse marks synchrony measures that were not able to indicate high significant effects below p-values of 1\%. Spike detection threshold factor was 5. }  
\label{Fig:Result_Experimental}
\end{figure}

In addition to the evaluation with \textit{in silico} manipulated data, all synchrony measures were applied to experimental data recorded from cortical neurons by \ac{MEA} chips with and without the application of 10\,$\mu$M \ac{BIC}. Fig.~\ref{Fig:Result_Experimental}~(a) shows the absolute synchrony values for each synchrony measure and for all different cell cultures ($N=5$), before and after the application of \ac{BIC}. 
Generally all measures showed a significant synchrony increase due to the \ac{BIC} application with p-values of 5\% and below, where A-SPIKE-synchronization, A-ISI-distance, and A-SPIKE-distance failed to reach the high significance level of 1\%. 
 
\begin{figure}[h]
%\nopagenumber
%\renewcommand{\baselinestretch}{1.0}
\hfill
\begin{center}
\includegraphics[width=.6\textwidth]{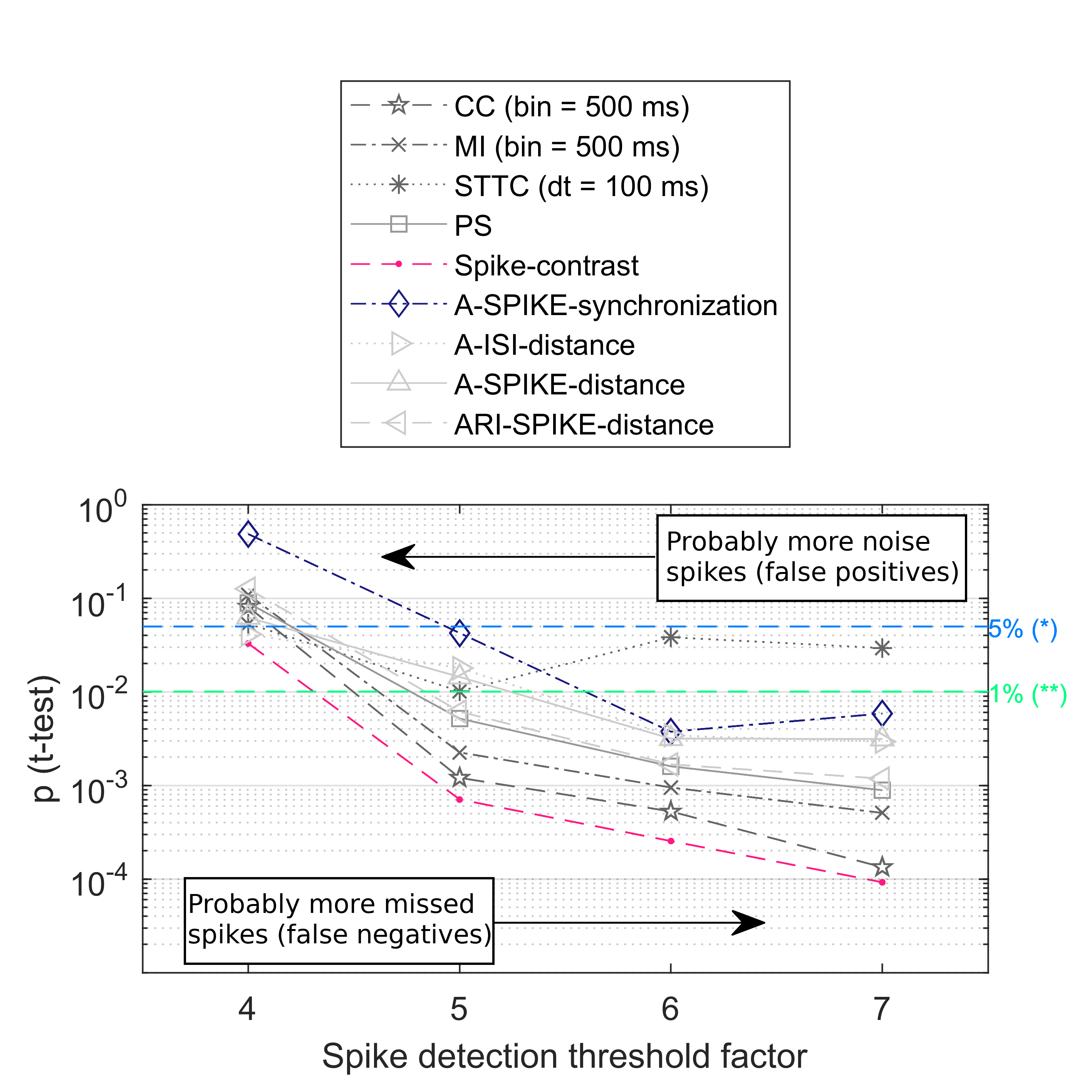}
\end{center}
\caption[]{\textbf{Influence of the spike detection threshold on the statistical significance of the experimental data.} Synchrony and statistical significance of the experimental data were analyzed like in Fig.~\ref{Fig:Result_Experimental} applying different thresholds to detect spikes from the recorded signals. Assumably, the higher the spike detection threshold factor, the less noise were detected but also more real spikes were missed. The lower the p-value, the better the synchrony measure's ability to capture the effect of \ac{BIC}.  }  
\label{Fig:Result_Experimental_th}
\end{figure}

In order to alter the level of false positive and false negative spikes in the experimental data, different spike detection threshold factors (4 to 7) were applied. Where low thresholds are likely to correspond to additional false spikes (false positive) and high thresholds to missed spikes (false negative). 
Increasing the threshold factor is comparable with deleting spikes from the spike train as already done in the \textit{in silico} manipulated data evaluation (Section~\ref{sec:simulation}). 
For the \textit{in silico} manipulated data, the synchrony measures \ac{STTC} and A-SPIKE-synchronization were most sensitive to deleted spikes. This behavior was also evident within the experimental data (Fig.~\ref{Fig:Result_Experimental_th}). As the spike detection threshold factor increased, \ac{STTC} and A-Spike-synchronization lose their statistical significance (above 5\% level), while all others remained stable indicating statistical significance (below 5\% level). Spike-contrast yielded the highest statistical significance across all tested threshold factors, and was the only measures that still indicated a statistical significance (at 5\% level) at the smallest threshold factor of 4, where a high level of noise spikes is assumable.

%%%%%%%%%%%%%%%%%%%%%%%%%%%%%%%%%%%%%%%
% Conclusion
%%%%%%%%%%%%%%%%%%%%%%%%%%%%%%%%%%%%%%%
\section{Discussion and Conclusion}
In this work we compared different spike train synchrony measures regarding their robustness to false positive and false negative spikes in epileptiform signals. Such robustness is particularly relevant for experimental data with error prone spike detection, especially at low signal-to-noise ratios. Representative synchrony measures were chosen from different categories such as time-scale dependent (\ac{CC}, \ac{MI}, \ac{STTC}), or time-scale independent (Spike-contrast, \ac{PS}, A-SPIKE-synchronization, A-ISI-distance, A-SPIKE-distance, ARI-SPIKE-distance).
In order to perform a comparison, we proposed a procedure based on a set of \textit{in silico} manipulated spike trains with defined manipulation of features.
Two data sets were generated based on experimental data by adding spikes, representing noise (false positive), and deleting spikes, representing missed spikes (false negative). 
Synchrony of the experimental data was increased by applying \ac{BIC} to cortical \textit{in vitro} networks. The code and data used in this work are publicly available (see Section~\ref{Section_SynchronyMeasures}). \newline

For the \textit{in silico} manipulated data, the results showed that Spike-contrast was most robust to added spikes and very robust to deleted spikes. All other measures showed either comparable high robustness for added or deleted spikes, but not for both cases. Furthermore, \ac{CC}, \ac{MI}, and Spike-contrast work with binnend spike trains, which explains their robustness to deleted spikes as long as bursts (=many spikes occuring within a small time period) are still present in the data (see Fig.~\ref{Fig:Result_Hybrid}, (b~2)). A binned spike train will almost not change if only some spikes are deleted from a burst and if the bin size and burst duration are similar. As Spike-contrast automatically adapts its bin-size to the data, it is preferable over \ac{CC} and \ac{MI} for exploratory studies, where the time-scale is not known beforehand. 

For the experimental data, all synchrony measures captured the synchrony increase mediated by \ac{BIC} in a statistical significant way (below a p-value of 5\%). However, there were differences in performance as the synchrony measures \ac{CC}, \ac{MI}, \ac{PS}, Spike-contrast, and ARI-SPIKE-distance yielded values leading to a high statistical significance below p-values of 1\%. Note that for \ac{CC} the proposed bin size of 500~ms was used, but smaller bin sizes of around 300~ms would have overperformed Spike-contrast (Fig.~\ref{Fig:ParameterChoice}). 
Referring to our assumption ``the lower the p-value, the better the synchrony measure'' defined in Section \ref{sec:Statistics}, it must be taken into account that this assumption is controversial since the actual synchrony increase (=ground truth) is not known for the experimental data. Thus, a synchrony measure that overestimates the synchrony increase mediated by \ac{BIC} would incorrectly lead to a low p-value. \newline

For small sample sizes, as in our experiment ($N=5$), parameter-free statistics like the Wilcoxon signed-rank test are generally used to avoid assumptions about population distribution. Application of the Wilcoxon signed-rank test to our data yielded almost identical p-values for all synchrony measures (data not shown). In contrast, the t-test used in this work resulted in different p-values for each synchrony measure. So, it also depends on the choice of statistical test whether the choice of synchrony measures affects the final results.

Considering the results of the \textit{in silico} manipulated data, the measures \ac{CC}, \ac{MI}, Spike-contrast, and ARI-SPIKE-distance were most robust to deleted spikes (=false negative spikes). As mentioned above, these synchrony measures also showed the best performance in the experimental data. This suggests that robustness to false negative spikes correlates with the ability to quantify synchrony changes in experimental data. If so, it would also imply that the experimental data used in this work were more affected by false negative spikes, than by false positive spikes. 
In other words, more spikes were missed than noise was misinterpreted as spikes. This behavior could be used to draw conclusions about the quality of an experimental spike train from the rank of the p-values of different synchrony measures.

Like already mentioned, there were some synchrony measures, whose results of the \textit{in silico} manipulated data and experimental data did not correlate. This suggests, that robustness to false positive or false negative spikes is not the only factor to effectively capture synchrony changes in experimental data. Factors like robustness to temporal non-stationarity of the recording could also be different among the synchrony measures. For instance, \ac{PS}, A-SPIKE-distance, ARI-SPIKE-distance, A-ISI-distance, and A-SPIKE-synchronization are more adaptive to changes in time scales inside a spike train as they dynamically define their time scales considering nearby spikes. In contrast, \ac{CC}, \ac{MI}, \ac{STTC}, and Spike-contrast use equally spaced time scales along the entire spike train duration.  

Overall, for our specific data set, Spike-contrast was the only time-scale independent measure being robust to noise (false positive spikes) as well as missed spikes (false negative spikes). This desirable performance was confirmed by the ability of the Spike-contrast measure to detect biochemically induced synchrony with a high significance, even for different spike detection threshold factors. It should be mentioned that the measures A-SPIKE-synchronization, A-ISI-distance, A-SPIKE-distance, and ARI-SPIKE-distance are able to produce a synchrony profile over time, which complements the synchrony profile over time-scale of Spike-contrast.  

Summarizing, we suggest to include the Spike-contrast synchrony measure into synchrony studies of epileptiform experimental neuronal data sets in addition to established synchrony measures.

%%%%%%%%%%%%%%%%%%%%%%%%%%%%%%%%%%%%%%%
% Acknowledgment
%%%%%%%%%%%%%%%%%%%%%%%%%%%%%%%%%%%%%%%
\subsection{Acknowledgments}

F.A.R. acknowledge CNPq (grant 305940/2010-4), Fapesp (grant 2012/51301-8 and 2013/26416-9) and NAP eScience - PRP - USP for financial support. L.F.C. thanks CNPq (grant 307085/2018-0) and FAPESP (grant 15/22308-2) for sponsorship. G.F.A. acknowledges FAPESP (grant 2012/25219-2). T.P. acknowledges FAPESP (grant 2016/23827-6). C.C.H. acknowledges FAPESP (grant 18/09125-4). C.T. and M.C. acknowledges Bayerisches Staatsministerium f\"ur Bildung und Kultus, Wissenschaft und Kunst for financial support in the frame of ZEWIS, BAYLAT, the BMBF (grant 03FH061PX3), and INeuTox (grant 13FH516IX6). C.N. would like to acknowledge support from the Studienstiftung des Deutschen Volkes.

%%%%%%%%%%%%%%%%%%%%%%%%%%%%%%%%%%%%%%%
% Appendix
%%%%%%%%%%%%%%%%%%%%%%%%%%%%%%%%%%%%%%%
\section{Appendix}
\subsection{Synchrony measure description} \label{Appendix1}

\paragraph{Mutual Information} \label{sec:mi} 

\begin{figure}[h]
\hfill
\begin{center}
	\includegraphics[]{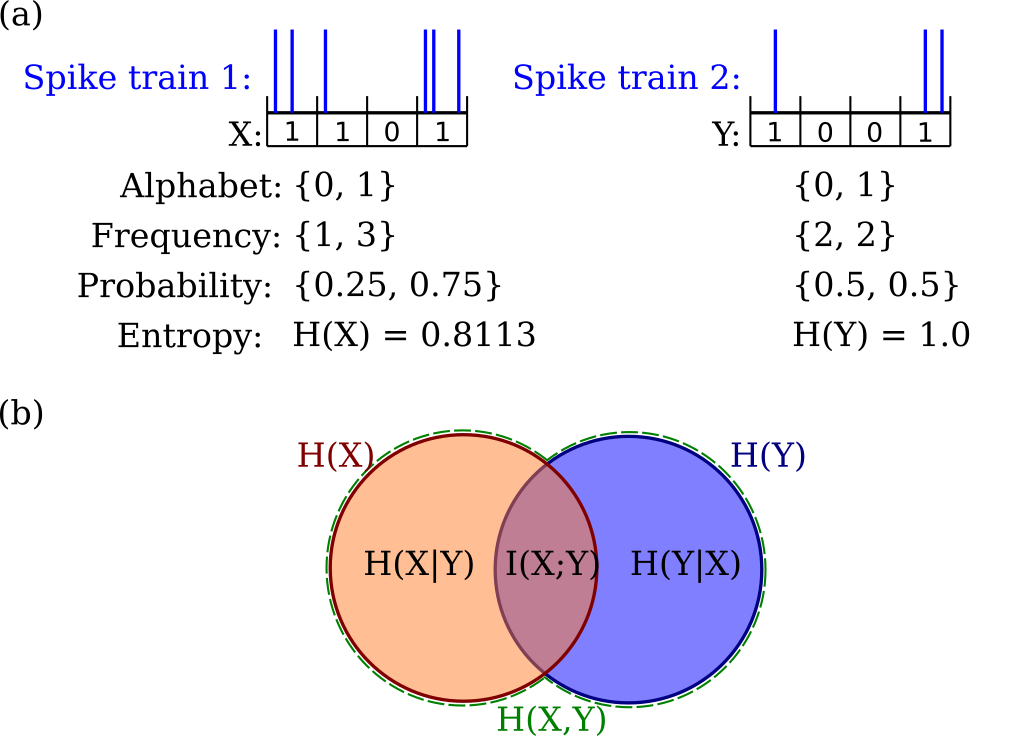}
\end{center}
	\caption{{\bf Example entropy calculation of two spike trains and illustration of the \acf{MI} measure.} (a) Two different spike trains are binned in a binary way resulting in X and Y. Each value is considered  a character building the alphabet of the signal. The probability of each character is obtained by dividing its incidence by the length of the signal. The entropy of each signal (H(X), H(Y)) is calculated using Eq.~\eqref{eq:entrop} with the probabilities given above. (b) Visualization of the mutual information I(X;Y) using the joint entropy H(X,Y), the entropies H(X), H(Y), and conditional entropies H(X$|$Y), H(Y$|$X) like calculated in Eq.~\eqref{eq:mut}. The mutual information value I(X;Y) increases with increasing synchrony between signal X and Y.}
	\label{Fig:MI}
\end{figure}

\ac{MI} measures how two random variables $X$ and $Y$ are related~\citep{Cover012} and is based on the concept of entropy, which is a fundamental concept in information theory~\citep{shannon27mathematical, Cover012}. 
The Shannon entropy~\citep{Cover012} of a random variable $X$ is defined as
\begin{equation}
 H(X) = - \sum_i p_x(i) \log(p_x(i)),
 \label{eq:entrop}
\end{equation}
where $p_x(i) = P(X = i)$ and the $\log$ function is taken to the base~$2$ (see Fig.~\ref{Fig:MI}~a for an example calculation). $H(X)$ measures the uncertainty about a random variable $X$. The conditional entropy quantifies the information necessary to describe the random variable $Y$ given that the information about $X$ is known~\citep{Cover012}. Formally, it is defined by
\begin{equation}
\begin{aligned}
H(Y|X)\ &\equiv \sum_{x\in X}\,p(x)\,H(Y|X=x)\\
&{=}\sum_{x\in X} \left(p(x)\sum_{y\in Y}\,p(y|x)\,\log\, \frac{1}{p(y|x)}\right)\\
&=-\sum_{x\in X}\sum_{y\in Y}\,p(x,y)\,\log\,p(y|x)\\
&=-\sum_{x\in X, y\in Y}p(x,y)\log\,p(y|x)\\
&=\sum_{x\in X, y\in Y}p(x,y)\log \frac {p(x)} {p(x,y)}. \\
\end{aligned}
\end{equation}

\ac{MI} is defined as
\begin{equation}
 I(X;Y) = \sum_{y \in Y} \sum_{x \in X} 
                 p(x,y) \log{ \left(\frac{p(x,y)}{p(x)\,p(y)}
                              \right) }. 
\label{eq:mi}
\end{equation}

Additionally, the \ac{MI} can be expressed in terms of the entropy and the conditional entropy of random variables $X$ and $Y$ as
\begin{equation}
\begin{aligned}
I(X;Y) & {} = H(X) - H(X|Y) \\ 
& {} = H(Y) - H(Y|X) \\ 
& {} = H(X) + H(Y) - H(X,Y) \\
& {} = H(X,Y) - H(X|Y) - H(Y|X).
\end{aligned}
\label{eq:mut}
\end{equation}
Aside from the analytic expressions, it is possible to interpret those quantities graphically, as depicted in Fig.~\ref{Fig:MI}b. \ac{MI} is more general than the correlation coefficient and quantifies how the joint distribution $p(x,y)$ is similar to the products of marginal distributions $p(x)p(y)$~\citep{Cover012}. Compared to the cross-correlation measure, mutual information also captures non-linear dependencies.

For comparison, the value of mutual information needs to be normalized. Many possible approaches have been proposed~\citep{Cover012}, such as
\begin{equation}
 M(X;Y) = \frac{I(X;Y)}{\min\left[ H(X),H(Y)\right]},
 \label{eq:mut1}
\end{equation}
because $I(X;Y) \leq \min\left[ H(X),H(Y)\right]$ and, consequently, $0 \leq M(X;Y) \leq 1$. This formulation has been used before in the context of neuronal signal analysis~\citep{Bettencourt2007, PhysRevLett.100.238701, Ham2010}. Another possible normalization is the so-called symmetric uncertainty~\citep{WittenFrank}, defined as
\begin{equation}
 M^*(X;Y) = 2\frac{I(X;Y)}{H(X) + H(Y)},
 \label{eq:mut2}
\end{equation}
where $0 \leq M^*(X;Y) \leq 1$. In this work the latter normalization were used as it performed better than the first one when applied to the experimental data (data not shown).

In order to estimate the \ac{MI} between two spike trains, both spike trains were transformed into binary binned signals. A spike train $i$ is binnend as
\begin{equation}
x_i(t)  = \begin{cases} 
			\text{$1$, if spike train $i$ shows at least one spike in time interval $t$ to $t + \Delta t$.} \\
			\text{$0$, if spike train $i$ shows no spike in time interval $t$ to $t + \Delta t$.}
			\end{cases}
	\label{eq:binning}
\end{equation}
using a bin size $\Delta t =$ 500\,ms and $t = 0\Delta t, 1\Delta t, 2\Delta t, ...$ (also see Fig.~\ref{Fig:MI}~a). The choice of the bin size of 500\,ms is justified in Section~\ref{sec:ParameterChoice}). 

\paragraph{Phase Synchronization} \label{sec:ps}
Since synchronization processes are related with rhythm adjustment, it is natural to introduce the concept of phase of an oscillator, a quantity that increases by $2\pi$ within an oscillation cycle and determines unambiguously the state of a periodic oscillator~\citep{Pikovsky03}. 
For instance, consider a harmonic oscillator described by the variable $x(t) = A\sin(\omega_0 t + \phi_0)$. In this case, $\omega_0$ denotes the angular frequency, which is related to the oscillation period $\omega_0 = 2\pi /T$, $A$ is the amplitude of oscillation, and the quantity $\phi (t)=\omega_0 t + \phi_0$ is the phase of this oscillator. 
Two or more oscillators are synchronized when they present the same phase evolution~\citep{Pikovsky03}. 

\begin{figure}[h]
\hfill
\begin{center}
\includegraphics[]{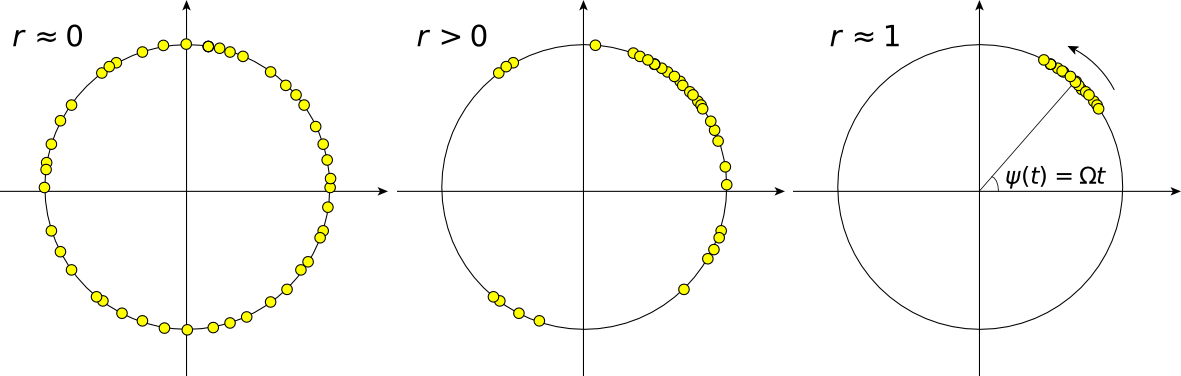}
\end{center}
	\caption{\textbf{Example for distribution of phase vectors $e^{i\phi (t)}$ in complex plane.} (a) Randomly distributed phase vectors over $[0,2\pi]$, implying $r \approx 0$; in other words the oscillators are completely asynchronous. (b) Regime of partial synchronization; phase vectors of some oscillators are grouped in a synchronous cluster equivalent to $r > 0$. (c) Strongly synchronized state, where all oscillators group into a single synchronous cluster rotating with average frequency $\Omega$, thus $r\approx 1$.}
	\label{Fig:coherence_figure}
\end{figure}

The measurement of the synchronization level of self-sustained oscillators can be done by considering the phases as rotating points in the unit cycle of the complex plane~\citep{Pikovsky03,strogatz2000kuramoto}. If an oscillator has a phase $\phi (t) $, then its trajectory in the complex plane is described by the vector $e^{i\phi (t)}$. For instance, if two oscillators have phases $\phi_1 (t) = \phi_2(t) = \phi (t)$, then they will have the same trajectory in the complex plane and thus the modulus of the resultant vector will be $|e^{i\phi_1(t)} + e^{i\phi_2(t)}|/2 = 1$, meaning that the oscillators are perfectly synchronized. Supposing now a population of $N$ interacting phase oscillators whose phases are described by the variable $\phi_i(t)$, $i=1,...,N$. The synchronization order parameter is defined as~\citep{strogatz2000kuramoto}
\begin{equation}
r e^{i\psi(t)} = \frac{1}{N}\sum_{j=1}^N e^{i\phi_j(t)},
\label{eq:order_parameter}
\end{equation}
where $\psi (t)$ is the average phase of the system at time $t$. When $r \approx 0$, the phases are distributed uniformly over $[0,2\pi]$ corresponding to the asynchronous state. When $r \approx 1$ the phases rotate together corresponding to fully synchronized state (Fig.~\ref{Fig:coherence_figure}).

Neurons can also be considered as self-sustained oscillators~\citep{Arenas08:PR} and are often modeled as integrate-and-fire oscillators, where the rhythmic quantity is the rate at which spikes are fired. The synchronization process in this case is the adjustment of the spiking patterns, i.e., if two interacting integrate-and-fire oscillators discharge their spikes jointly, then they are synchronized. Therefore, in order to quantify the synchronization between interacting integrate-and-fire oscillators, the phases associated to the respective spike signals have to be defined.
 
\begin{figure}[h]
\hfill
\begin{center}
	\includegraphics[]{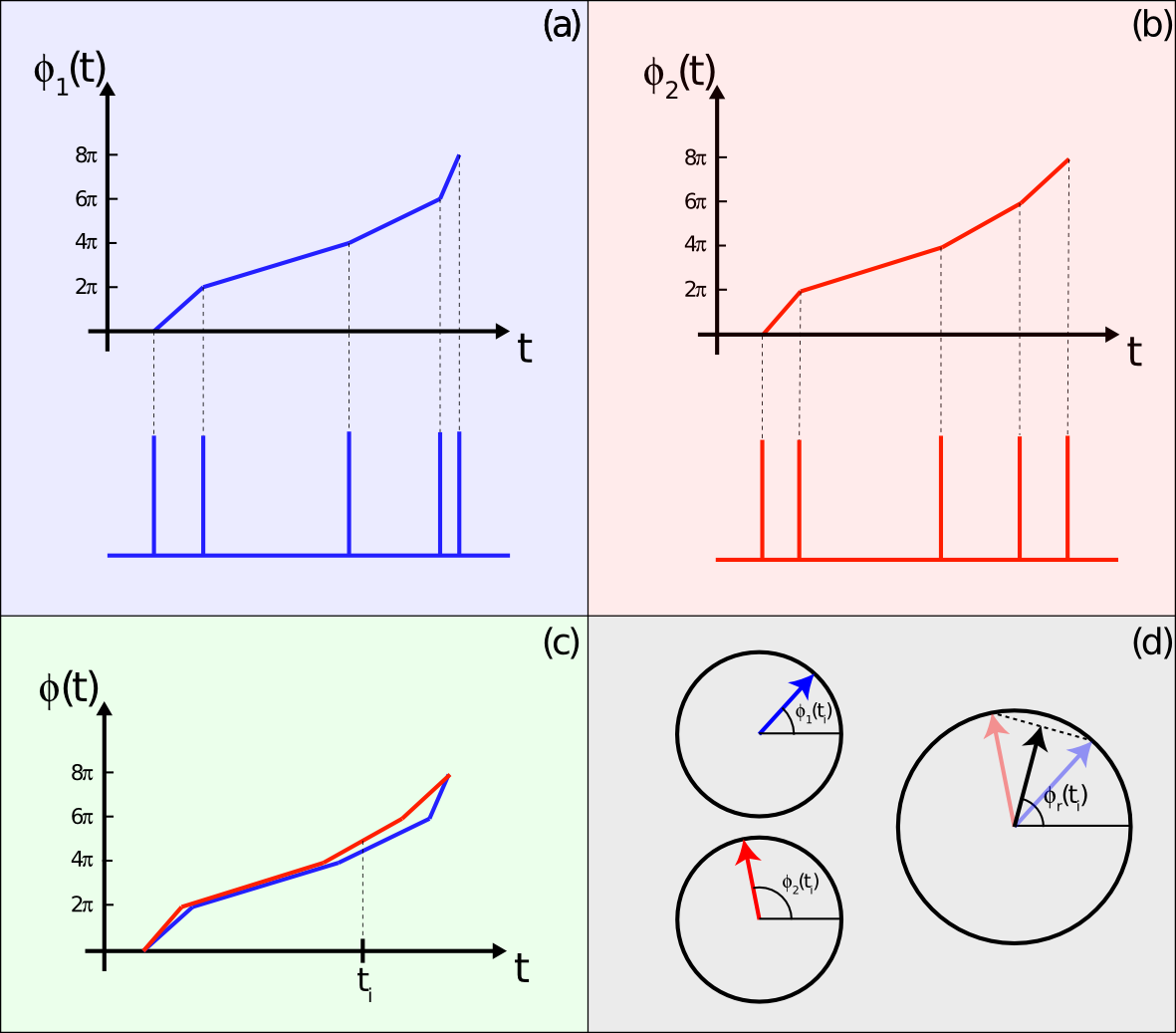}
\end{center}
	\caption{\textbf{Example of phase vector construction.} (a) $\phi_1 (t)$ and (b) $\phi_2 (t)$ are calculated using the respective spike signals according to Equation~\eqref{eq:phases}. (c) Superposition of temporal evolution of phases and (d) phase vectors $e^{i\phi_1 (t_i)}$,  $e^{i\phi_2 (t_i)}$ as well as the resultant vector at time $t=t_i$, $r\cdot e^{i\phi_r (t_i)} = (e^{i\phi_1 (t_i)} + e^{i\phi_2 (t_i)})/2 $.}
	\label{Fig:coherence_figure_example}
\end{figure}

The time series recorded by an electrode can be considered as a sequence of point events taking place at time $t_k$, with $k=1,2,...,N_{\mathrm{spikes}}$. The time interval between two spikes can be treated as a complete cycle. In this case, the phase increase during this time interval is exactly $2\pi$. Hence, the values of $\phi(t_k)=2\pi k $ are assigned to the times $t_k$, and for an arbitrary instant of time ($t_k < t < t_{k+1}$), the phase is considered as a linear interpolation between these values~\citep{Pikovsky03,neiman1998stochastic,pikovsky1997phase,tass1998detection,hu2000phase}, as
\begin{equation}
\phi(t) = 2\pi \frac{t - \tau_k}{\tau_{k+1}  - \tau_{k}  }+ 2\pi k.
\label{eq:phases}
\end{equation}
This method can be applied to any process containing time series of spikes and it is widely used in the study of synchronization in neuronal dynamics \citep{Pikovsky03}. Fig.~\ref{Fig:coherence_figure_example} exemplifies the calculation of phases $\phi (t)$ of two spike signals and their respective sum of the phase vectors $e^{i\phi (t)}$ in the complex plane.  

For a system composed of $N$ oscillators, the instant phase synchronization is quantified
by the order parameter defined in Eq.~\eqref{eq:order_parameter}. However, to quantify the level of synchronization of the neurons, we consider the average over the recorded time
\begin{equation}
\bar{r}  = \left\langle \left| \frac{1}{N}\sum_{j=1}^N e^{i\phi_j(t)} \right|  \right\rangle_t, 
\label{eq:order_parameter_average}
\end{equation}
where $\left\langle \cdots \right\rangle_t$ stands for the temporal average and $N$ stands for the number of electrodes used to compute the order parameter.

%%%%%%%%%%%%%%%%%%%%%%%%%%%%%%%%%%%%%%%
% Author contributions
%%%%%%%%%%%%%%%%%%%%%%%%%%%%%%%%%%%%%%%
\section{Author contributions}

% M.C. Manuel Ciba
% C.N. Christoph Nick
% R.B. Robert Bestel
% T.P. Thomas Peron
% G.F.A. Guilherme Ferraz de Arruda
% C.C.H Comin Cesar Henrique 
% L.F.C. Luciano da Fonoura Costa 
% F.A.R. Francisco Aparecido Rodrigues
% C.T. Christiane Thielemann

M.C.: Conception, design, generation, and analysis of \textit{in silico} manipulated data. Application of the synchrony measures STTC, Spike-contrast, A-SPIKE-synchronization, A-ISI-distance, A-SPIKE-distance, ARI-SPIKE-distance to experimental data. Statistical tests and spike detection with varying thresholds. Preparation of figure 1 to 6. Draft of the manuscript.
R.B.: Initial conception and design of synchrony measure comparison using experimental data.
C.N.: Initial conception and design of synchrony measure comparison using experimental data. Performed cell experiments and spike detection of raw data.
G.F.A.: Implementation and application of the synchrony measure MI to experimental data and MI description in appendix.
T.P.: Implementation and application of the synchrony measure PS to experimental data and PS description in appendix. Preparation of figure 7 and 8.
C.C.H.: Implementation and application of the synchrony measures CC to experimental data. 
L.F.C.: Helped in the initial conception, discussions and revision.
F.A.R.: Conception and design of research.
C.T.: Conception and design of research.
ALL: Interpreted and discussed results of experiments. Edited and revised manuscript. Approved final version of manuscript.

%%%%%%%%%%%%%%%%%%%%%%%%%%%%%%%%%%%%%%%
% Bibliography
%%%%%%%%%%%%%%%%%%%%%%%%%%%%%%%%%%%%%%%
%\bibliographystyle{elsarticle-harv} 

\end{document}